\documentclass[runningheads]{llncs}
\usepackage{graphicx}
\usepackage{url}
\begin{document}
\title{ILoveEye: Eyeliner Makeup Guidance System with Eye Shape Features}

\author{Hange Wang \and
Haoran Xie \and
Kazunori Miyata}
\authorrunning{H. Wang et al.}
\titlerunning{ILoveEye}
\institute{Japan Advanced Institute of Science and Technology, Ishikawa 9231292, JAPAN \\
\email{xie@jaist.ac.jp}}

\maketitle             % typeset the header of the contribution
\begin{abstract}
Drawing eyeliner is not an easy task for whom lacks experience in eye makeup. Everyone has a unique pair of eyes, so they need to draw eyeliner in a style that suits their eyes. We proposed ILoveEye, an interactive system that supports eye-makeup novices to draw natural and suitable eyeliner. The proposed system analyzes the shape of the user's eyes and classifies the eye types from camera frame. The system can recommend the eyeliner style to the user based on the designed recommendation rules. Then, the system can generate the original patterns corresponding to the eyeliner style, and the user can draw the eyeliner while observing the real-time makeup guidance. The user evaluation experiments are conducted to verify that the proposed ILoveEye system can help some users to draw reasonable eyeliner based on the specific eye shapes and improve their eye makeup skills\footnote{This is a pre-peer review, pre-print version of this article. The final authenticated version is available online at proceedings of HCII 2022(International Conference on Human-Computer Interaction)}.

\end{abstract}
\section{Introduction}

Makeup is a daily skill like other creative activities such as painting and sculpting, which requires the users to master skillful techniques of makeup and possess a high-level understanding of makeup based on facial features. Eye makeup plays an important role in the daily facial makeup. However, it is still a challenging issue for common users with few makeup experiences to achieve satisfactory outcome. Usually, the users often seek makeup tutorials from video sites or social medias, such as Youtube and TikTok. These video tutorials commonly do not match personal facial features, which have significant relationship with the strategies of eye makeup. In this work, we focus on the guidance system for eyeliner makeup which is the basic facial makeup in our daily lives. 

About makeup guidance systems, it is a common approach to provide the corresponding makeup suggestions based on the user's facial features and show the anticipated makeup effect on the face in real time. We found that it is beneficial to provide visual recommendation makeup styles and visual makeup tutorials for whom has few knowledge and experience of makeup. This work aims to provide intuitive and effective eyeliner makeup guidance in real time. 

 In this paper, we propose, ILoveEye, an interactive rule-based makeup system based on the analysis of user’s eye shape features for less-experienced users as shown in Figure 1. To obtain the feature points of eyes, we utilize state-of-the-art deep learning based facial recognition models in this work. To classify the features of eye shapes, we conduct eye shape classification and extract the features of the eye contour. After analyzed the typical eye shapes including almond, round, downturned, and close-set eyes, we propose a rule-based classification model to obtain the labels of the eye shape features with three feature values of the eyes: eye aspect ratio, outer corner orientation, and eye distance. For the makeup guidance, we investigate rules of typical eyeliner styles according to the relationship between eye shape and eyeliner styles. In terms of the eye feature points obtained from the proposed system, we adopt the classification rules to match current feature points of user's eye. The proposed system can reproduce the eyeliner styles and display in real-time as guidance to support the user to complete the eyeliner makeup. 

\begin{figure}[t]
\includegraphics[width=\textwidth]{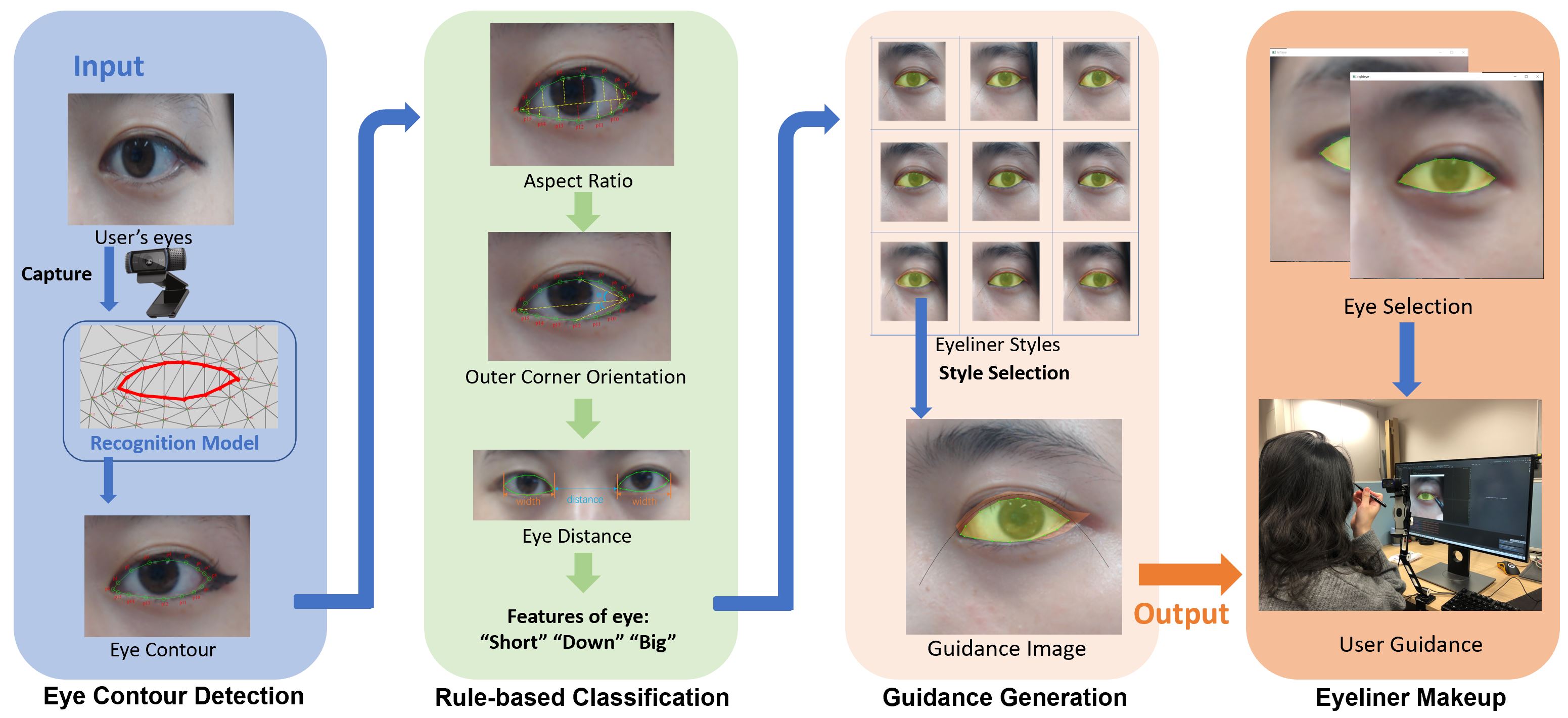}
\caption{The proposed makeup guidance framework of ILoveEye.} 
\label{fig:framework}
\end{figure}

We designed and conducted an evaluation experiment to verify the feasibility of the proposed ILoveEye system. In our user study, we invited participants to join the experiments, who were less-experienced in facial makeup with few skills of drawing eyeliner. To clarify the relationship between system evaluation and user’s skill background, we designed a questionnaire which were conducted after experienced the proposed system. From the questionnaire results, we found that the proposed system could effectively classify the eye shape and support the user to draw eyeliner interactively. In addition, we analyzed the Pearson correlation coefficients between the user’s makeup background and system evaluation. The analysis results indicated a significant and positive correlation between the frequency of wearing makeup and the skill improvement after using the system. We conclude that the users who normally wear makeup are acquired with experience and knowledge of makeup, thus they could draw better eyeliner after learning the exact contours of their eye shape features using the proposed system.

\section{Related Work}
\subsection{Eye Features}
The analysis of eye features plays an important role in eye makeup system. Bhat et al. proposed a method for detecting eye contours by using active shape model~\cite{Bhat08}. An efficient method for detecting eye contours was proposed in real-time with the dataset of the eye contours~\cite{Fuhl17}. For shape classification of eyes, Alzahrani et al. summarized the characteristics of eye shapes and designed eye shape classification rules~\cite{Alzahrani21}. In this work, we originally design the classification rules by collecting relevant information and confirm the effectiveness of our classification rules by conducting evaluation study.  

\subsection{Guidance System}
Along the development of augmented reality and image processing technologies, the guidance systems for supporting our daily activities have been explored extensively. The deep learning based human mesh modeling was used to compare the user's postures and target ones for core training \cite{training2019}. To help take a good selfie, a voice guidance system was proposed with crowdsourcing evaluation of photo aesthetic \cite{Fang2018SelfieGS}. The spatial augmented reality has been used for calligraphy practice \cite{he2020}, golf training \cite{golf19} and dance support \cite{dance21}. To meet the user's intentions in design activities, the user sketches were utilized for domino arrangement    \cite{peng2020sketch2domino} and lunch box decoration \cite{sketchmehow}. In this work, we especially focus on the guidance system for eye makeup using augmented reality approach. 

\subsection{Makeup Guidance}
With the development of facial recognition technologies, it becomes feasible to provide suitable makeup guidance based on facial features. Chiocchia et al. developed a mobile application to help users select cosmetics and showed users makeup tutorials~\cite{Chiocchia21}. Face features can be obtained for makeup image synthesis with different makeup references~\cite{guo09}. The face dataset was constructed to analyze the facial attractiveness for recommend system \cite{liu13}. A rule-based system was proposed to automatically classify the faces and suggest the makeup approach \cite{Alashkar17}. iMake proposed a novel eye-makeup design system to extract the color and shape features~\cite{Nishimura14}. A mirror system utilizing augmented reality (AR) was proposed to capture and analyze the facial features in real-time \cite{Rahman10}. An interactive mobile application was proposed for makeup tutorials \cite{Almeida15}. A makeup instruction with AR was proposed to provide special makeup tools~\cite{Treepong18}. In this work, we aim to provide the real-time guidance to support users for drawing better eyeliner and improving makeup skills.

\section{System Overview}

Figure~\ref{fig:framework} shows the overall workflow of the proposed ILoveEye system which consists of three main parts with deep learning based face detection. First, the proposed system detects eyes using the normal web camera and obtains the geometric information of the eye contours. Second, we design a rule-based model using geometric features to classify eye shapes. Finally, the proposed system generates the eyeliner makeup guidance based on the previously analyzed classification results displayed on the screen in real-time. The proposed ILoveEye system enables users to observe the guidance patterns on the screen while drawing suitable eyeliners.

\subsection{Eye Detection}
\begin{figure}[htb]
\includegraphics[width=\textwidth]{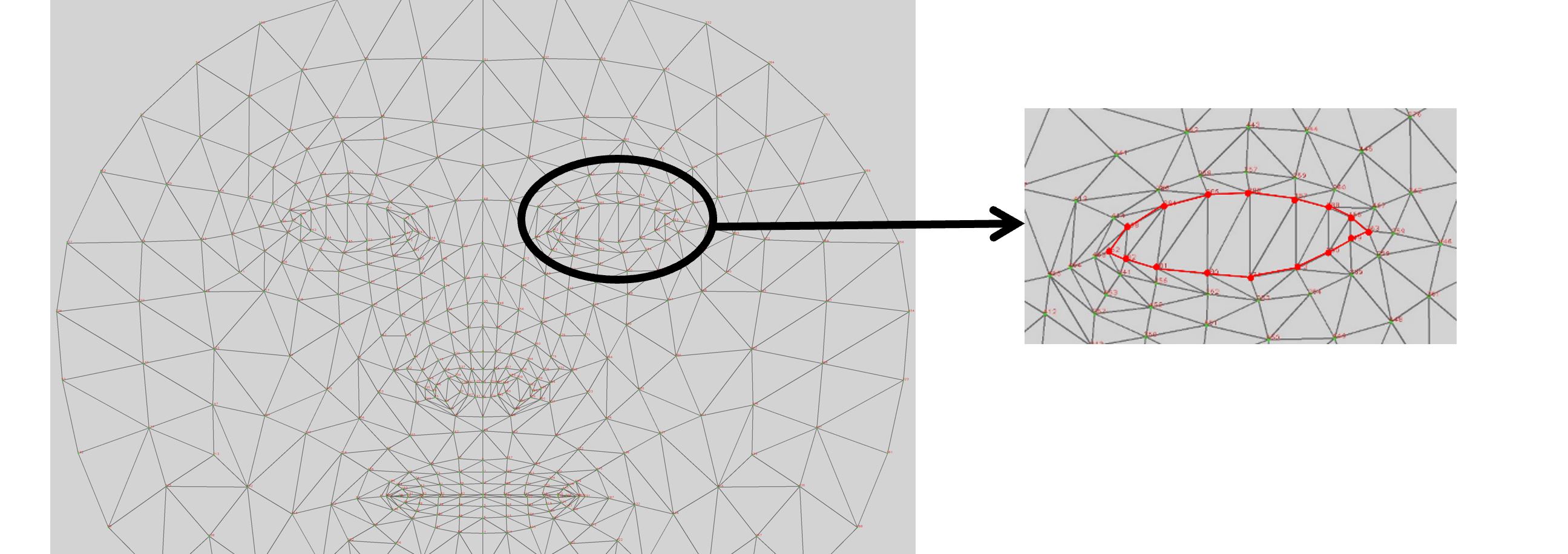}
\caption{Eye contour landmarks in human face mesh.} 
\label{fig:mesh}
\end{figure}

About eye shape detection, we adopted the open-source deep learning based face detector,  MediaPipe \cite{MediaPipe}, which contains a pre-trained face recognition model with high efficiency and accuracy. The detection output is a face mesh map as shown in Figure~\ref{fig:mesh} with 486 landmarks. From this face mesh, 16 feature points are selected to describe the eye contour. Through our practical testing, we found that this detector can effectively identify the shape of the eye contour with high accuracy to meet our research target.

To implement the eye detection module, we first invoke the connected camera (60 fps in our prototype). To get the shape of the eyes accurately, the proposed system asks the user to save the mesh map of the current frame. The proposed system selects the landmarks of the eye parts from the saved  mesh map. We found that the camera angle may affect the detection results, so we asked the user to sit at a proper distance and allow the user to adjust the camera angle until the satisfied eye contour is achieved.

\subsection{System Workflow}

\begin{figure}[htb]
\centering
\includegraphics[width=\textwidth]{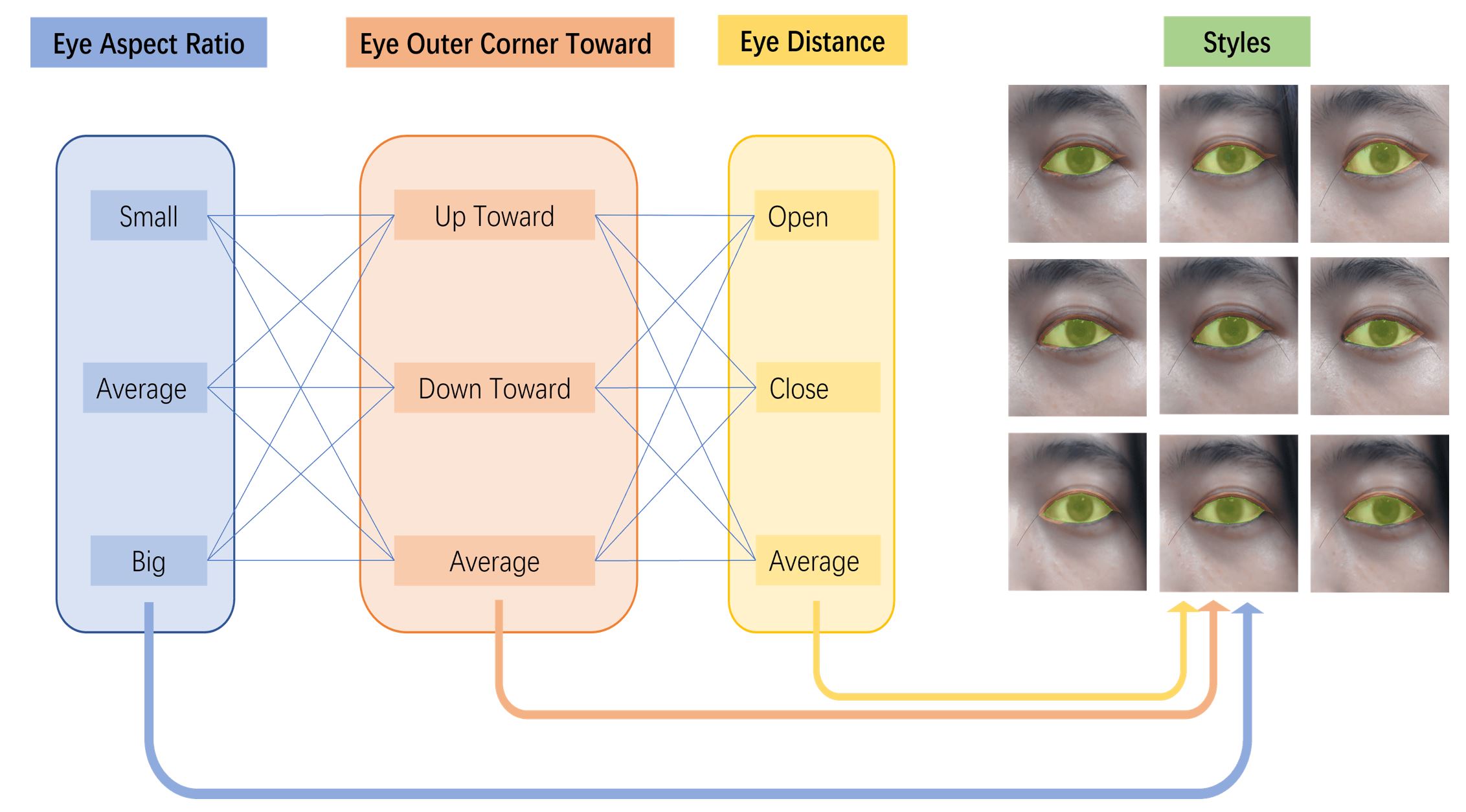}
\caption{Workflow of the recommendation function.}
\label{fig:system}
\end{figure}

After analyzed the eye features, the proposed system then determines the eyeliner styles in terms of eye aspect ratio, outer corner orientation, and eye distance as shown in Figure~\ref{fig:system}. If the eye size is labeled with ``small" or ``average", the system recommends Style-Basic. If the aspect ratio is labeled with ``big", the proposed system recommends Style-Basic with the lower eyeliner 
thickness. The system can analyze the orientation of the eye outer corner to recommend Style-Winged, Style-Drop and Style-Extend. The system can recommend lower eyeliner styles: Style-Inner and Style-Outer, depending on eye distances. Finally, the proposed system combines all recommended eyeliner styles to generate a polygon as a mask, which is visualized in orange color for makeup guidance (Figure~\ref{fig:system}).

\subsection{Eye Features Analysis}
We reviewed different typical eye types, such as round eyes: larger and more circular; close-set eyes: less space between eyes; down-turned eyes: taper downward at the outer corner \cite{Alzahrani21}. We found that the eye shapes are related to the size of the eye, the angle of the eye outer corner, and the distance between the two eyes. Therefore, we set these three features as the classification conditions for determining eye types.

\begin{figure}[htb]
\centering
\includegraphics[width=\textwidth]{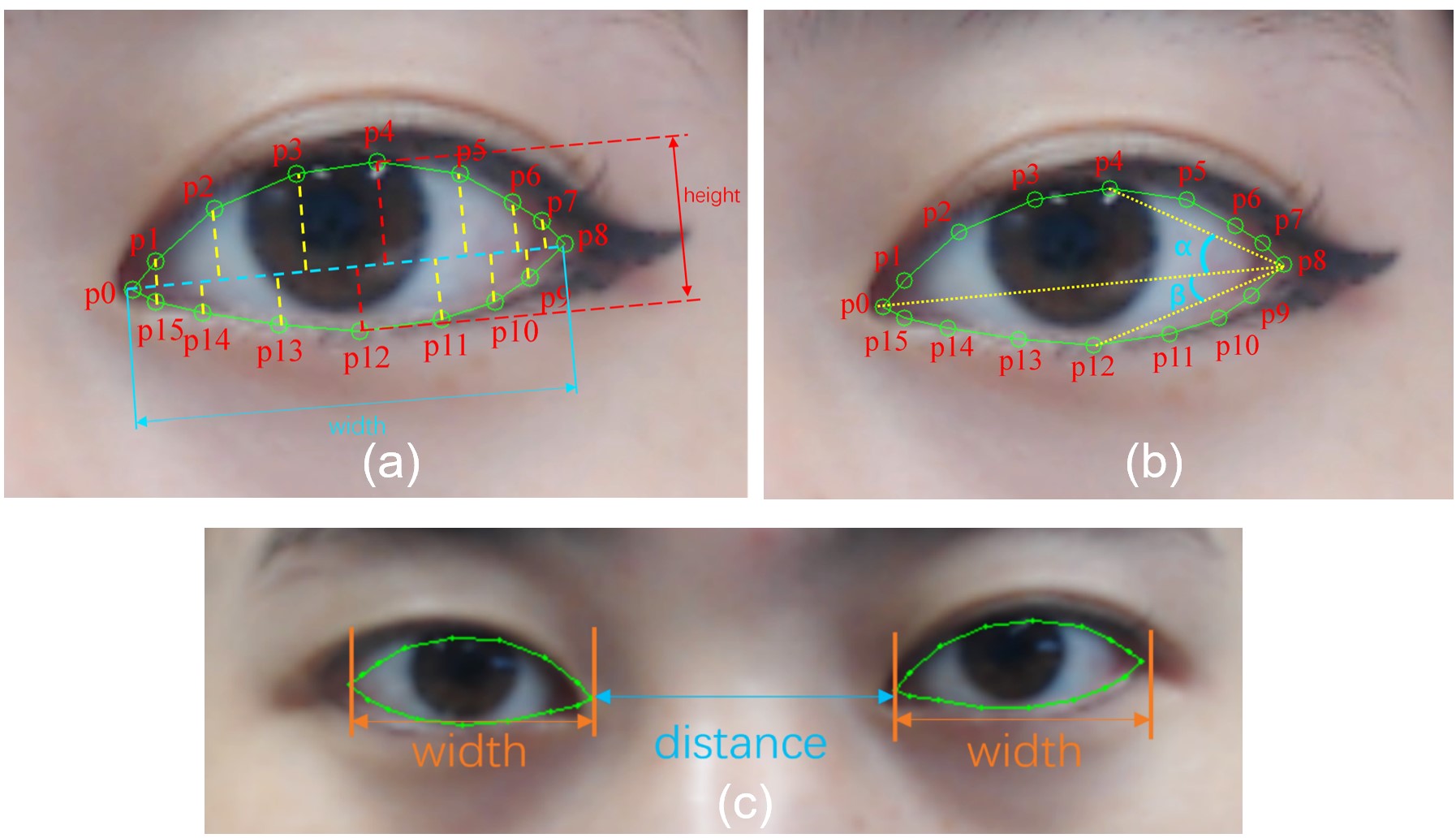}
\caption{Eye feature parameters: (a) width and height of eye; (b) outer corner orientation; (c) eye distance.}
\label{fig:parameter}
\end{figure}

\subsubsection{Aspect Ratio}
In the obtained eye contour landmarks (the right eye as an example), we define the leftmost and rightmost points $p0$ and $p8$ as the positions of the head and tail positions. We calculate the lengths of the perpendicular lines from $p1-p7$, and $p9-p15$ respectively, then find the maximum values of the upper part and lower part, and sum them as eye height  (Figure~\ref{fig:parameter}(a)). The length of $p0-p8$ line is considered as eye width. 
The ratio of eye width and height is calculated as aspect ratio $a$. We obtained a relatively reasonable range for eye styles: $a \in [2.75, 3.00]$ for average eye, small long eyes for $a>3.00$, big round eyes for $a<2.75$.

\subsubsection{Outer Corner Orientation}
We found that the angle of the eye's outer corner can be used as a reliable reference, for determining up-turned or down-turned eyes. As shown in Figure~\ref{fig:parameter}(b), $p4$ and $p12$ denote the highest and lowest points. Line $p0-p8$ divides the outer corner angle into two angles. The upper angle $\alpha$ is defined as included angle of lines $p4-p8$ and $p0-p8$, and lower angle $\beta$ with $p12-p8$. If $\alpha > \beta$, the outer corner is determined to be down toward and the eye will be labeled as down-turned eye. On the contrary, the eye will be labeled as upturned eye.

\subsubsection{Eye Distance}
We measure the distance between point most left and right points as eye-distance $D_e$. The average distance of two eyes $D_{avg}$ (Figure~\ref{fig:parameter}(c)) is calculated. If $D_e > 1.05 \times D_{avg}$, the eyes are considered as open-set type and labeled as “open”; If $D_e < 0.95 \times D_{avg}$, the eyes are considered as close-set type and labeled as “close”. Otherwise, the eyes are labeled as “average”. 

\section{Makeup Guidance}
\begin{figure}[t]
\centering
\includegraphics[width=0.8\textwidth]{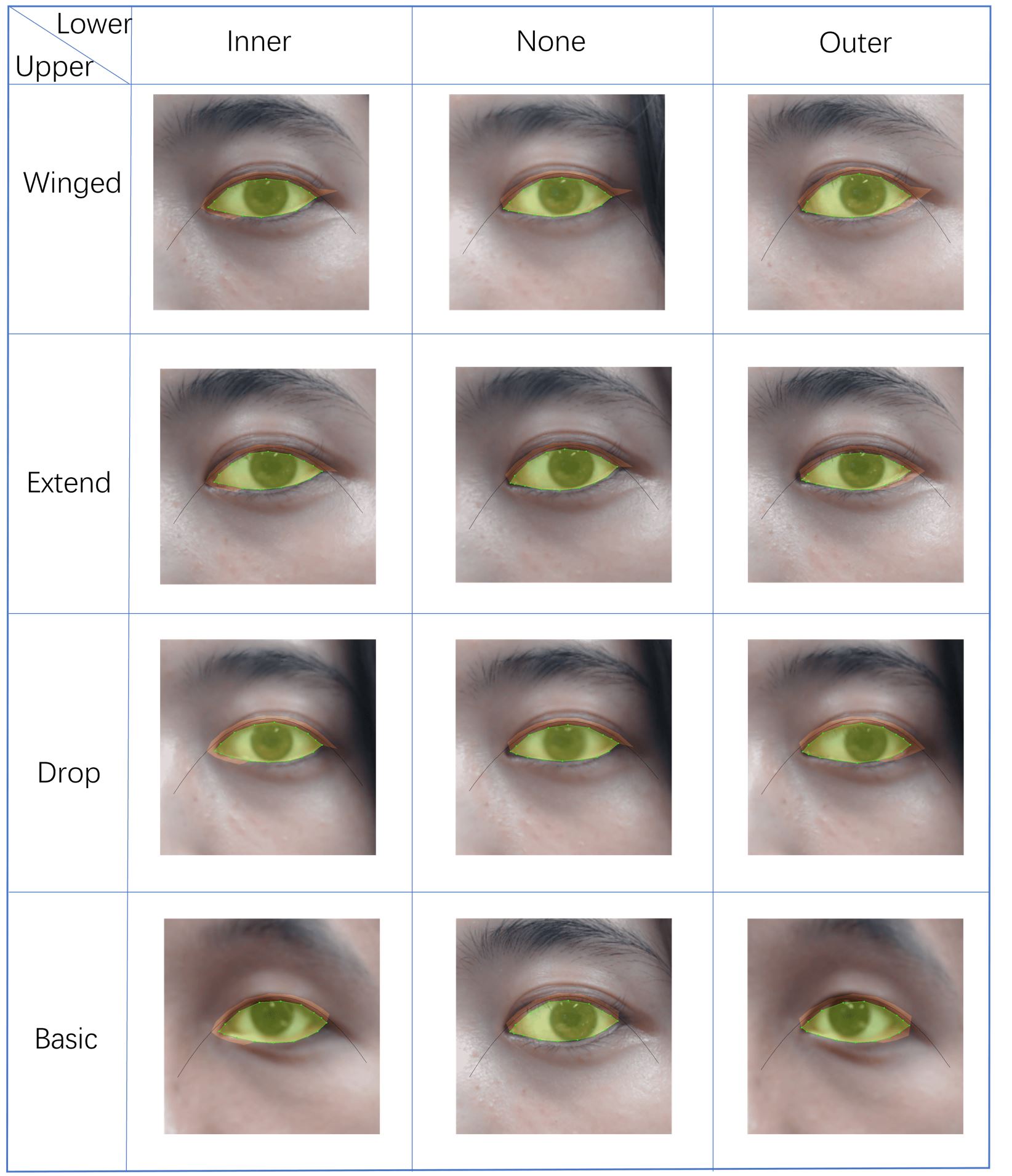}
\caption{Eyeliner styles defined in the proposed ILoveEye system.}
\label{fig:allstyle}
\end{figure}

Eye makeup uses eyeliner to make eyes look bigger and modify the eye shape \cite{Matsushita2015MeasurementOE}. Generally, there are two typical eyeliner types: outer and inner eyeliners. However, the inner eyeliner is applied in eyelashes roots which is difficult to detect by the proposed detection method. Therefore, we focus on the outer eyeliner in this work. We defined the typical styles of eyeliners related to eye shape features as shown in Figure~\ref{fig:allstyle}.

\subsection{Eyeliner Styles}
\subsubsection{Style-Basic} Style-Basic is a thick line along the upper eyelid that starts from the inner corner of the eye to the outer corner (Figure~\ref{fig:style1}(a)). In this work, a thick line is generated along with the shape of the upper eyelid. We calculate the mid-point for each line segment of the upper eyelid. Then, we set the perpendicular line of each line segments from mid-points to $p0’,p1’…p7’$, and the length of each perpendicular line is defined as eyeliner thickness $h$. By connecting the points of the upper eyelid, the closed polygon is the basic upper eyeliner style as Style-Basic.

\subsubsection{Style-Winged}
For the upturned-winged eyeliner pattern, we define an outermost point $E$. As shown in Figure~\ref{fig:style1}(b), we measure typical samples of winged eyeliner. The angle of the eyeliner tail is about 15 degrees for natural look, so we define the average angle between $E-p8$ and $p0-p8$ as 15 degrees. The length of the wing is set to be about 12\% of the eye-width. The wing length is also suitable for other wing eyeliner styles. The coordinates of $E$ point is calculated by trigonometric functions. By connecting the Style-Basic to the point $E$ to form a closed polygon, we get the upturned wing eyeliner as Style-Winged.

\subsubsection{Style-Drop}
As shown in Figure~\ref{fig:style1}(c), we get a droopy eye-line style by defining a point $E$. $p8-E$ is the extension of $p4-p8$, and the length of $p8-E$ is 12\% of eye-width in Style-Winged. The closed polygon formed by connecting Style-Basic to point $E$ is regarded as Style-Drop.

\subsubsection{Style-Extend}
As shown in Figure~\ref{fig:style1}(d), Style-Extend is the eyeliner style that the end of an eyeliner extends out horizontally. We first define the outermost point of the eye-liner as point $E$. Points $p0$, $p8$, and $E$ are on the same horizontal line, and the length of $E-p8$ is 12\% of eye-width. Finally, we connect Style-Basic with point $E$ to get Style-Extend.

\subsubsection{Basic Lower Eyeliner}
\begin{figure}[htb]
\centering
\includegraphics[width=\textwidth]{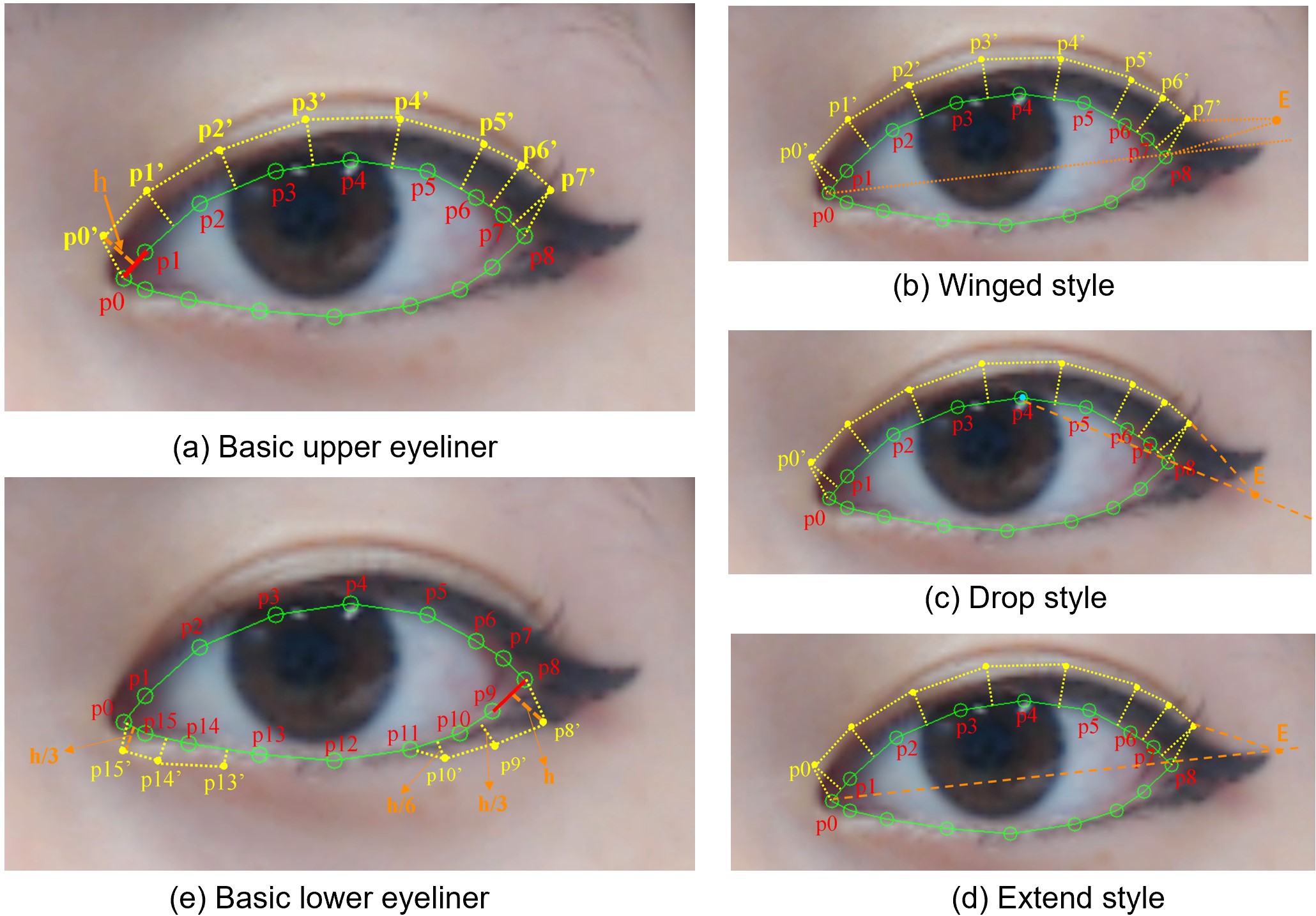}
\caption{Generation of different eyeliner styles.}
\label{fig:style1}
\end{figure}

The lower eyeliner is always thinner than the upper eyeliner and applied on corner sides.
For lower eyeliner in outer corner side (Figure~\ref{fig:style1}(e)), we connect the wing of upper eyeliner, and select $p8$, $p9$, $p10$, $p11$ to create a lower eyeliner style on the inner corner side. We decrease the value of $h$ from large to small in equal proportion to make the shape appearance from big to small. We connect the upper eyeliner and select $p13$, $p14$, $p15$, $p0$ to create a lower eyeliner style on the inner corner side. 
We found that the value of h as h/3 of the outer side can make this pattern look thinner than the outer side. We defined these two lower eyeliner styles as “Style-Inner” and “Style-Outer”.

\begin{figure}
\centering
\includegraphics[width=\textwidth]{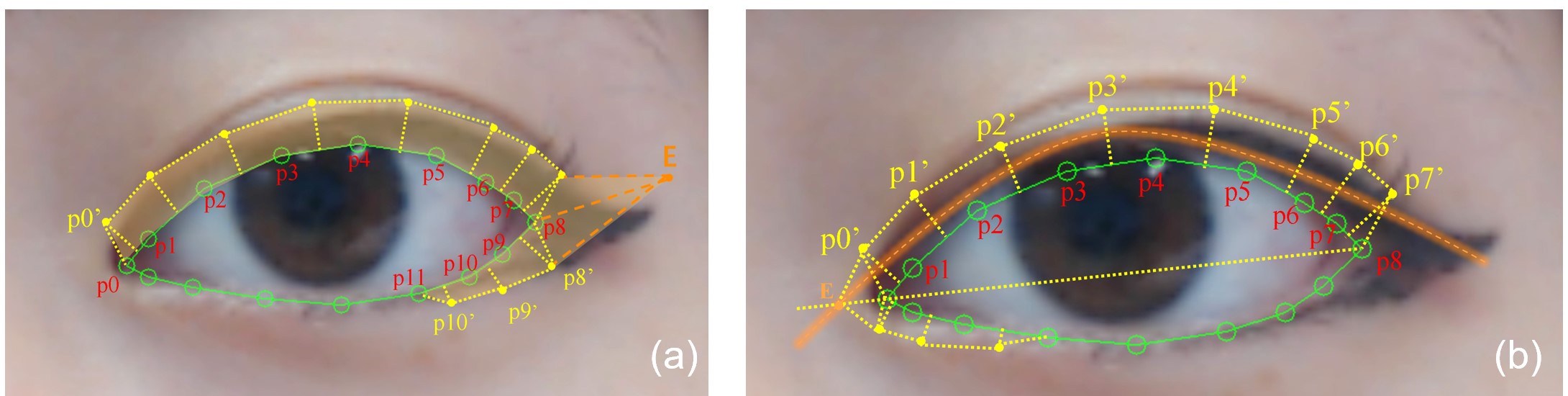}
\caption{Merging types of eyeliner styles.}
\label{fig:style2}
\end{figure}

\subsection{Merging Types}

For the lower eyeliner styles, the system selects the recommended style and merges it with upper eyeliner styles. In the case of merging Style-Outer with Style-Winged, the system connects point $E$ to lower 
$p8’$, and gets the closed polygon as the result for guidance as shown in Figure~\ref{fig:style2}(a).

In the case of merging Style-Inner and Style-Basic (Figure~\ref{fig:style2}(b)), the system first calculates a fitting curve by $p0-p8$ and $p0’-p7’$, then finds the cross-point $E$ with line $p0-p8$. We obtain a closed polygon by connecting the point $E$, Style-Basic and Style-Inner as shown in Figure~\ref{fig:style2}(b).

\subsection{User Interface}
\begin{figure}[htb]
\centering
\includegraphics[width=\textwidth]{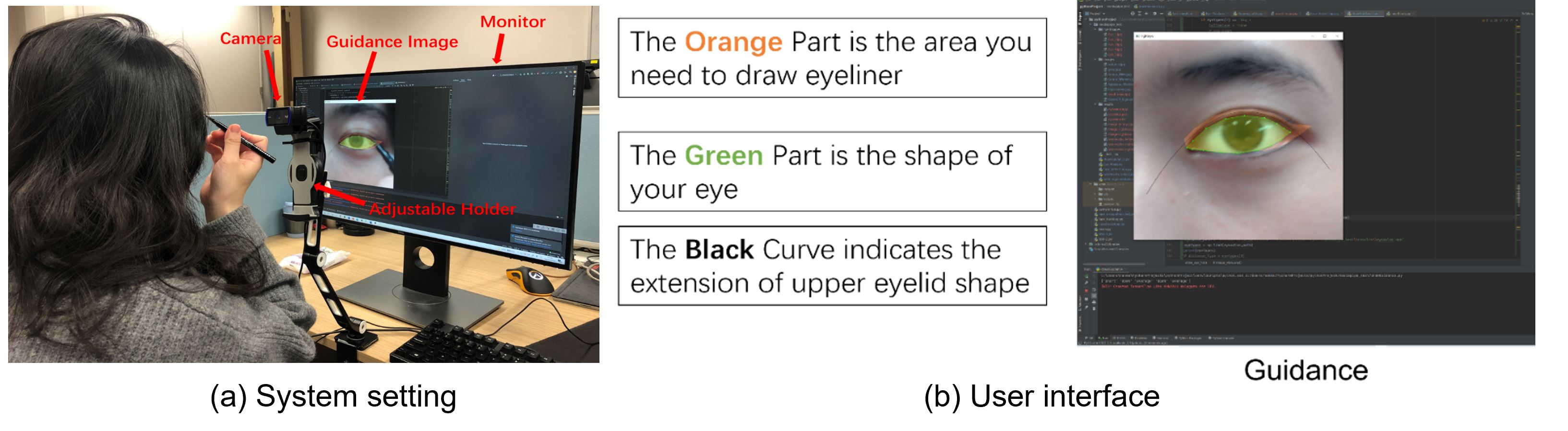}
\caption{Experiment Environment and user interface.}
\label{fig:ui}
\end{figure}

Figure~\ref{fig:ui} shows the user interface of the proposed ILoveEye system. The proposed system creates new windows of the left and right eyes separately and enlarges them to make the user observe the guidance and their eyes clearly. For the recognized eye contour, the system can display it that the user can find how well their eyes are recognized.

The user can select one window to save the feature information of the current frame through keyboard input. Because the feature information of both eyes is saved at the same time, the recognized eye shape may be in error with the real shape due to the shooting angle and lens distortion of the camera. Therefore, we asked the user save the eye features at the correct angle while looking straight to the camera.

\section{System Evaluation}

 The prototype system was implemented on a desktop computer with Windows 10, Intel i7-9700 CPU, GeForce RTX 2060 SUPER GPU. The proposed system was developed in Python with OpenCV ver4.5.2, MediaPipe ver0.8.6, and we set up Logitech HD Pro camera C920 on the table to capture user’s face.

\subsection{Experiment Procedure}
 We conduct a comparison study that participants are asked to draw eye makeup with and without the proposed system. After the eye-makeup is applied, we provided makeup remover for the participants to remove their makeup and start to draw makeup using the proposed system.
 We first introduced the system workflow and asked users to draw an eyeliner by referring to the makeup guidance in orange color. All participants were asked to look at the camera, the monitor can display the participant's face and independent images of the two eyes. The participants can adjust the angle of the camera until the satisfied eye recognition results were achieved. The analyzed eye features results were prompted to the participant in text. We provided participants with two eyeliners (one gel eyeliner and one liquid eyeliner) to draw eyeliner. When the participant completed the eye makeup, we counted the time cost and archived makeup images.

\subsection{Evaluation Method}
In the user study, we adopted both subjective evaluation and objective evaluation to verify the proposed ILoveEye system. Regarding to the subjective evaluation, we designed a questionnaire to confirm the user experience and the system usefulness. After collected the basic information of the participants (gender and age), we first investigated the users' knowledge about eye makeup. The question items of the questionnaire are listed as below.

\begin{itemize}
    \item Q1: Do you wear makeup?
    \item Q2: Do you watch makeup tutorials (videos, magazines, blogs)?
    \item Q3: How do you evaluate the level of your makeup skill?
    \item Q4: How do you evaluate the level of your eye-makeup skill?
    \item Q5: Do you draw eye-makeup by considering your eye shape?
    \item Q6: Do you consider this system describe your eye shape correctly? 
    \item Q7: Do you consider this system support you to draw eyeliner? 
    \item Q8: Do you consider this system help you to draw better eyeliner than by yourself 
    \item Q9: Do you think this system improve your drawing eyeliner skill?
\end{itemize}

All these questions were evaluated with 5-point Likert scale where 1 denotes strongly disagree and 5 for strongly agree. Finally, we collected the free feedback from the participants for any advice or improvement.  We invited 8 female graduate students, around 25-year-old. All participants were asked to join without eye makeup, and the few experience of eye makeup was required. 

\section{Results}

\begin{figure}[t]
\centering
\includegraphics[width=0.9\textwidth]{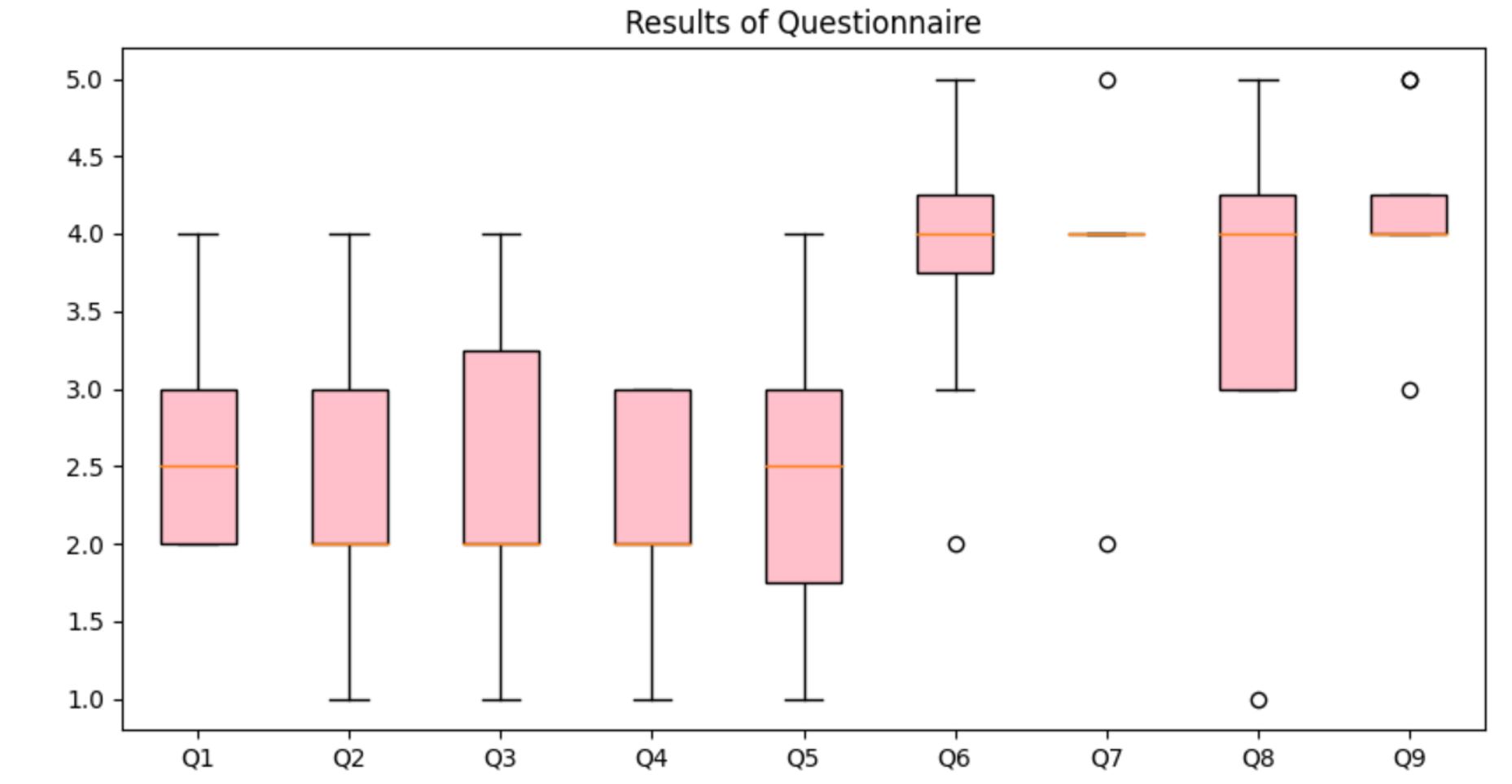}
\caption{Results of questionnaire in our user study.}
\label{fig:questionaire}
\end{figure}

\subsection{Subjective Evaluation}
The results of the questionnaire are shown in Figure~\ref{fig:questionaire}. We found that most participants did not wear makeup very often (Q1: average = 2.63) and were not confident in their makeup skills (Q3: average = 2.50; Q4: average = 2.25).  The average scores of questions from Q6 to Q9 were Q6: 3.88; Q7: 3.88; Q8: 3.63; Q9: 4.13. We found that the overall score of our system was above 3.0, which means our system can help users to draw better eye makeup and improve their makeup skills. We noticed that one participant rated lower scores on Q6, Q7, and Q8, and confirmed that it was difficult to satisfy experienced users well with our makeup guidance, who already had a well-tested routine for drawing eye makeup.
 
\begin{figure}[t]
\centering
\includegraphics[width=0.9\linewidth]{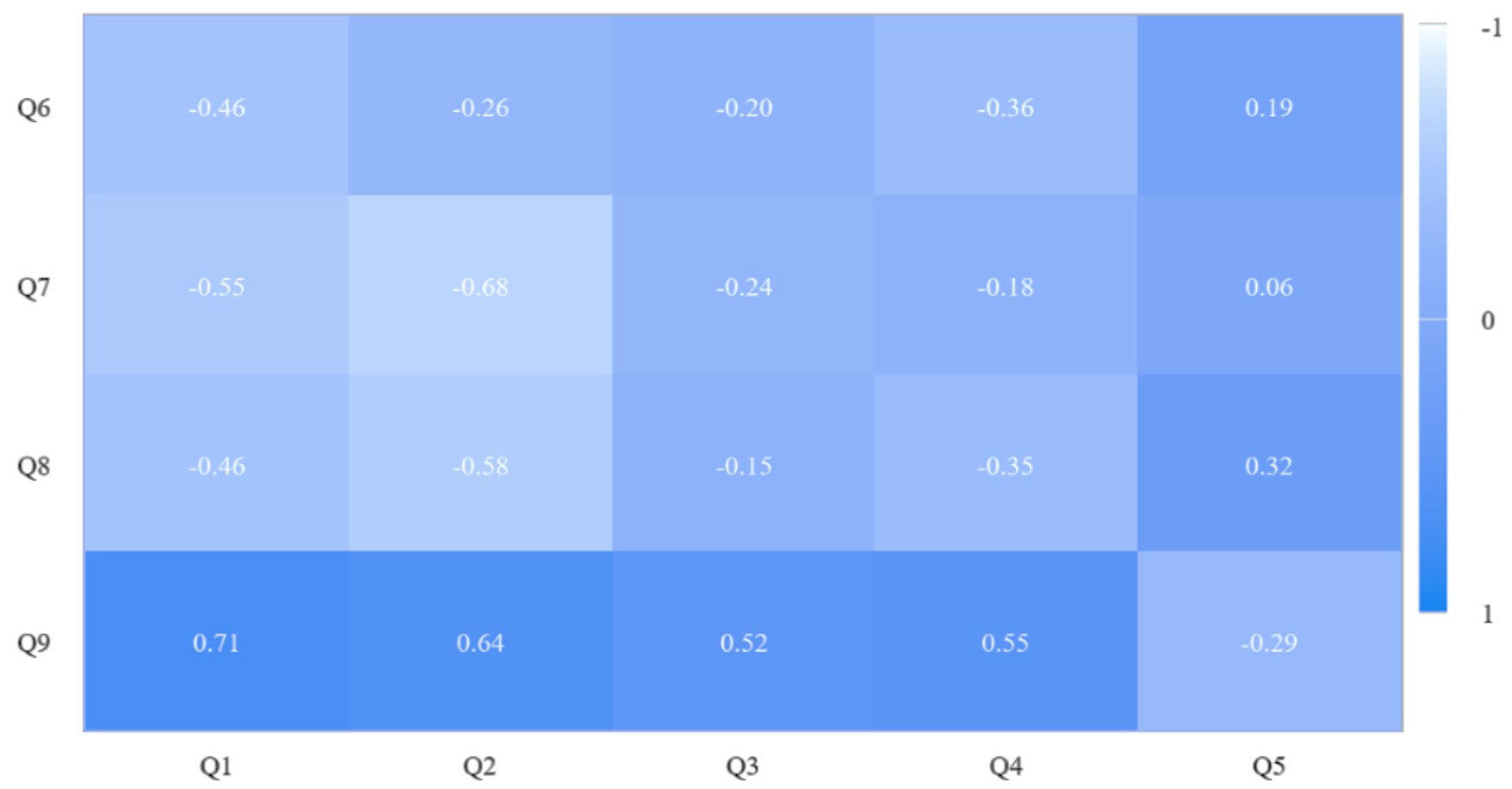}
\caption{Pearson coefficients for Q1-Q5 and Q6-Q9.}
\label{fig:Pearson}
\end{figure}

We analyzed the correlation between participants' background of makeup knowledge and evaluation of the system by the Pearson correlation coefficients for Q1-Q5 and Q6-Q9 as shown in Figure~\ref{fig:Pearson}. We found a significant and positive correlation between Q1 and Q9 (p = 0.71). This means that the participants who wear makeup frequently can feel the skill improvement with the proposed system. We observed that participants who normally wear makeup are acquired with some experience and knowledge of makeup. When they learn the exact contours of their eye features, it can help them to draw eyeliner afterward. In addition, one participant reported that a better experience for improving makeup skills than watching tutorial videos was obtained because the system could describe and show the eye features.

\begin{figure}[t]
\centering
\includegraphics[width=0.9\textwidth]{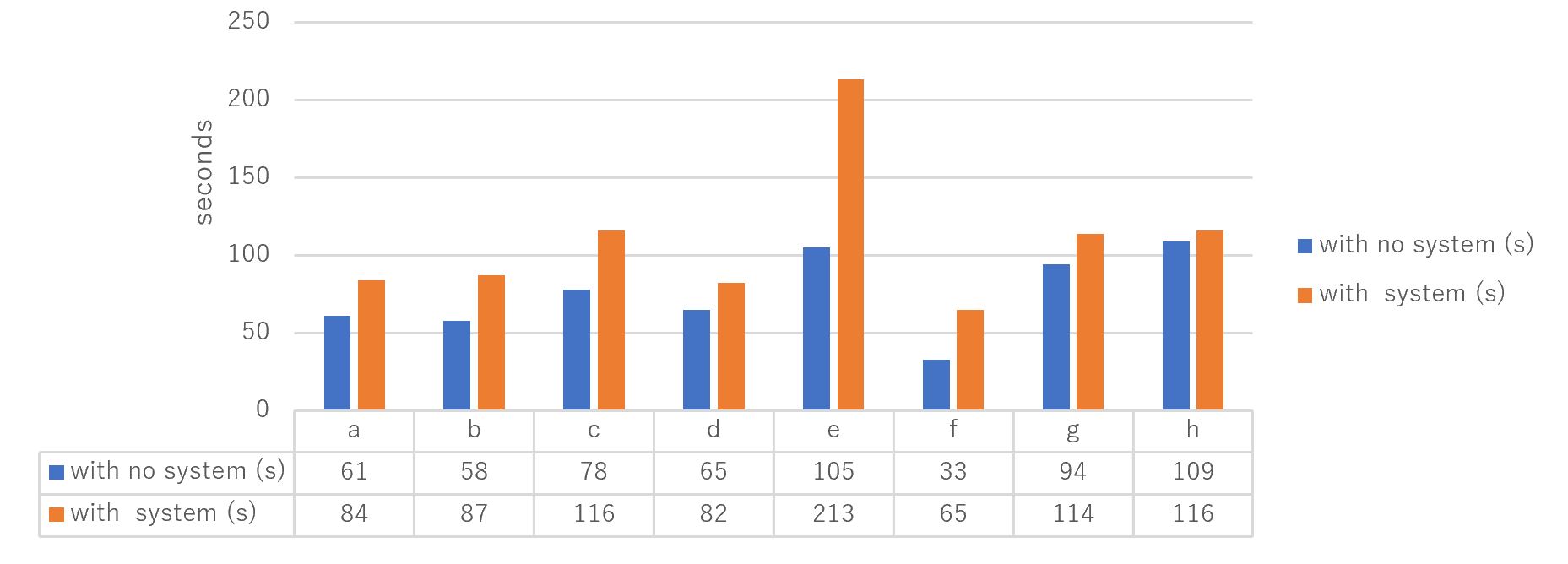}
\caption{Comparison of completion time in our user study.}
\label{fig:time}
\end{figure}

\subsection{Objective Evaluation}

In the objective evaluation, we compared the makeup completion time and makeup effect with and without the proposed system as shown in Figure~\ref{fig:time}. We found that after using our system, all users' completion time were increased. The users spent more time to find the location of the guidance pattern and consider how to reproduce the guidance pattern while using the system. In addition, we analyzed the correlation between completion time, and the questions in the questionnaire. We found a significant negative correlation between Q2 and completion time without system (p = -0.723), which means that the less frequently people watched makeup tutorials, the more time they spent on completing their eyeliner.

\begin{figure}[t]
\centering
\includegraphics[width=0.9\linewidth]{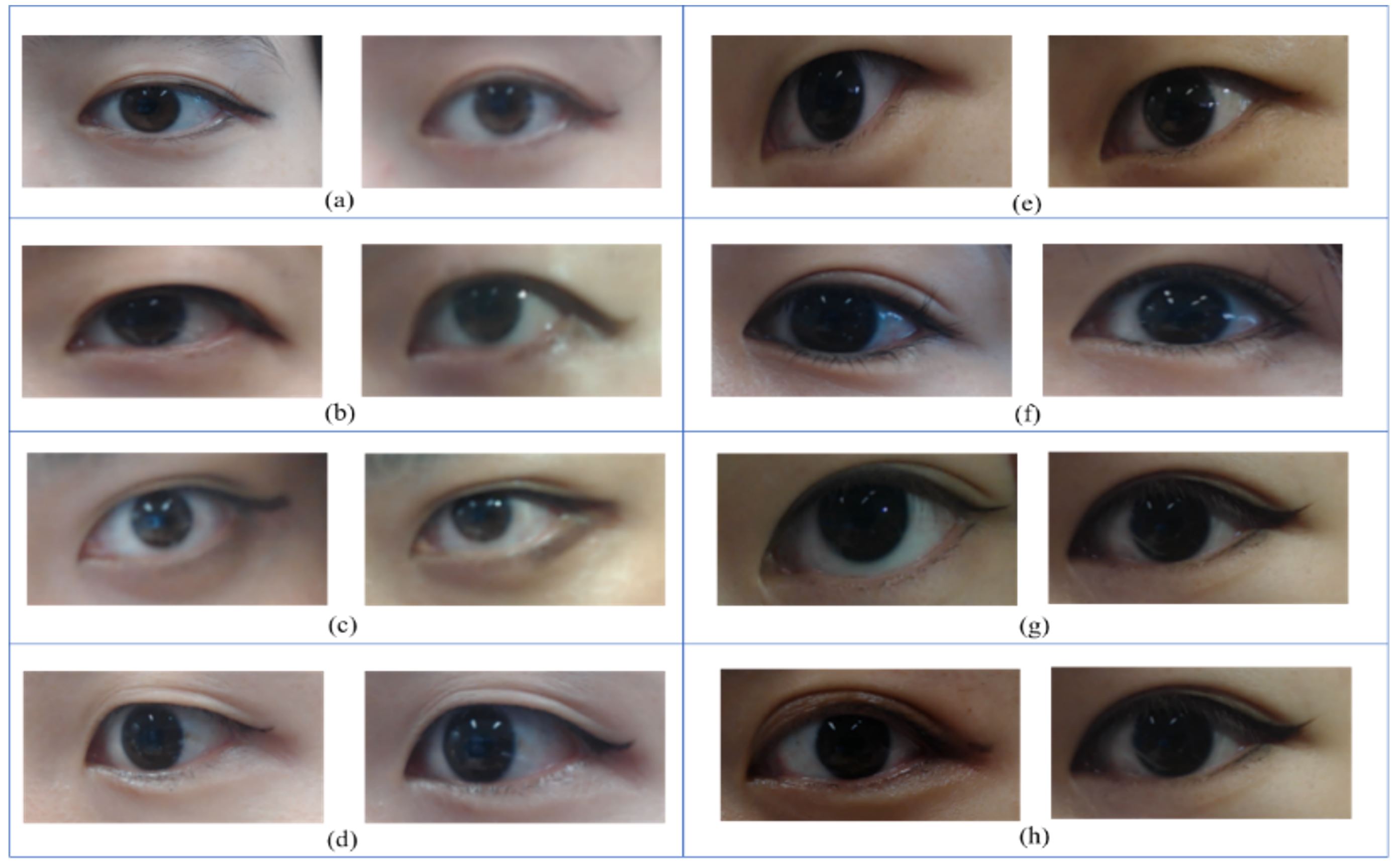}
\caption{Comparison of makeup effect by participants.}
\label{fig:re1}
\end{figure}

\begin{figure}[htb]
\centering
\includegraphics[width=0.9\textwidth]{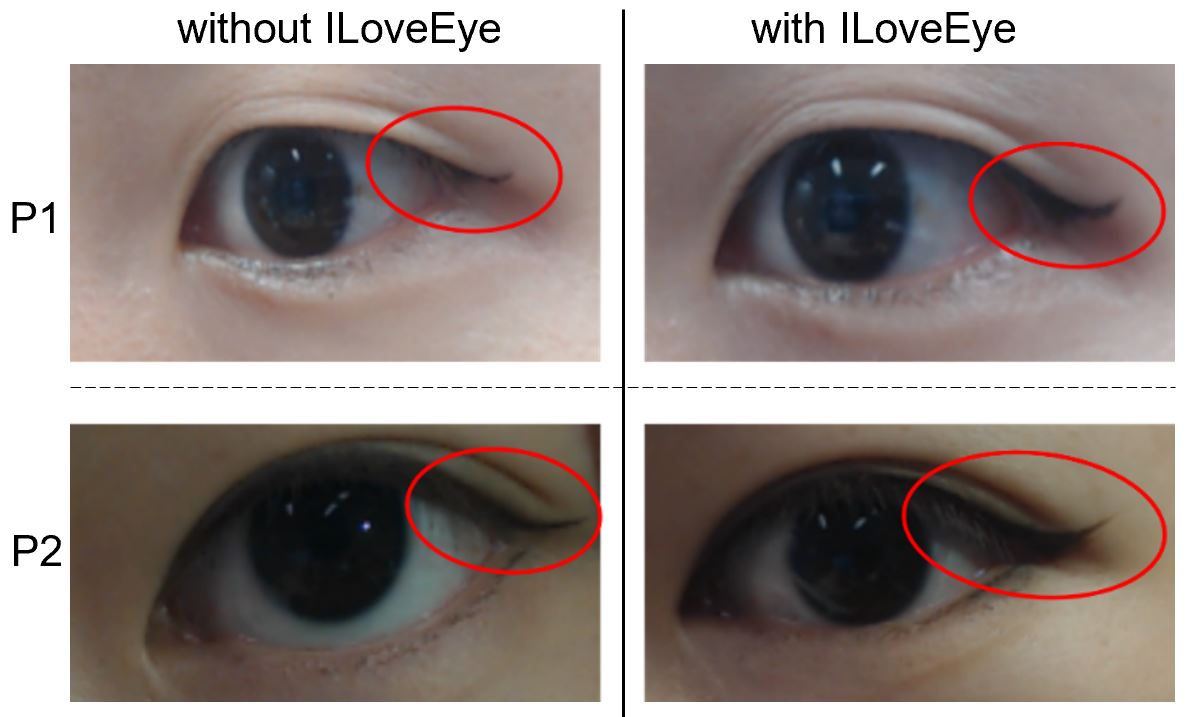}
\caption{ The difference of eyeliner's wing (Left without-system; Right: with-ILoveEye).}
\label{fig:re2}
\end{figure}

From the eye makeup results in both conditions as shown in Figure~\ref{fig:re1}, we found that the eye makeup which refers to the makeup guidance was generally thicker and more clearly than those without using the system. It is verified that the proposed makeup guidance  is easy to understand and could be reproduced by the users. The detailed comparison results are shown in Figure~\ref{fig:re2}. Observing the comparison results of winged parts, we found that the eyeliners drawn by the participants with the proposed system look thicker and much clearer. After reviewed the evaluation of system interactivity, it is verified that the makeup guidance provided by the proposed system was easy to understand and could be reproduced by users. Basically, these thicker and much clear eyeliners make users’ eyes became visually bigger.

\section{Discussion}
 From the interviews with participants and free comments in questionnaires, we found that the proposed system can effectively help users with less makeup experience to draw a better eyeliner that suits the feature of eye shape. Besides, from the collected participants feedback, we found that the proposed system could help users understand their eye features and improve their eye-makeup skills. However, most recommended eyeliner styles to the participants were Style-Wing, Style-Extend and Style-Outer, which means that they have similar eye shapes. All participants in this study were Chinese and Japanese, it would be better to verify the proposed system for more nationalities. For the eye recognition, it was difficult to achieve accurate recognition results due to camera angle, which was adjust the camera angle manually. Although the proposed system can capture the eye contour accurately, the detection of eyelids was very limited which means that the system may have difficulty for users when eyelids heavily affect the visual effect of the eyes.

In our user study, the users had to keep a distance from the monitor so that the nearsighted users may have difficulty to see the exact location following the guidance. The frame rates were sometimes low for the smoothness of observing the guidance. One participant reported that it would be helpful if the system could provide more recommended solutions for users to choose. For example, a scene choice function can be added for different usages such as the party, business and shopping. It would be promising to provide a face preview with the applied makeup styles.

\section{Conclusion}

In this work, we proposed ILoveEye, an eyeliner guidance system based on the analysis of eye shape. This system captures the user’s eyes by a camera and recognizes the eye contours for analyzing the eye shape. The proposed system can provide a recommended eyeliner guidance that users can draw eyeliners by referring to guidance. We utilized a deep learning based model to recognize the eye contour. We proposed the rule-based model to classify the eye shape features. We designed a user study to verify the effectiveness of the proposed system. From the user study, our system can effectively identify the user's eyes and recognize the contours of the eyes, and the user can follow the guidance to complete the eye makeup. The system provided an effective guidance function to help users draw better eye makeup. We found that users were satisfied with recommended guidance and the user with more makeup experience could obtain the improvement for their makeup skills.

To improve the user experience of our proposed system and help users draw eyeliner, there are some possible future works. In this study, the facial recognition model could not effectively identify the eyelids when the system recognized the user's eye contour, which may decrease the classification accuracy of the user's actual eye shape. To solve this issue, we plan to build a novel dataset containing images of eyes with labeled eyelid shapes and train a supervised learning model to optimize the effect of eye shape recognition. Furthermore, some participants reported that our proposed system provided few eyeliner styles, which may not be suitable for rare eye shapes, it is worth to designing more eyeliner styles as makeup guidance in the future to meet the user diversity. We think that the proposed ILoveEye system can be extended to improve other makeup skills such as eyebrows makeup.

\section*{Acknowledgement}
The authors thank the anonymous reviewers for their valuable comments. This project has been partially funded by JAIST Research Grant and JSPS KAKENHI grant JP20K19845, Japan.

\bibliographystyle{splncs04}
\bibliography{reference}
\end{document}